\documentclass[aip,apl,reprint,twocolumn]{revtex4-2}
\usepackage[utf8]{inputenc}
\usepackage[T1]{fontenc}
\usepackage{amsmath,amssymb}
\usepackage{float}
\usepackage{graphicx}
\usepackage{lineno}
\usepackage{blindtext}
\usepackage{xcolor}
\usepackage{tikz}
\usepackage{siunitx}
\usepackage{gensymb} 
\usepackage{todonotes}

\begin{document}
\title{Dark Matter Searches with Levitated Sensors}
\author{Eva Kilian}
\email{eva.kilian.18@ucl.ac.uk}
\author{Markus Rademacher}
\author{Jonathan M. H. Gosling}
\author{Julian H. Iacoponi}
\author{Fiona Alder}
\affiliation{%
 Department of Physics \& Astronomy, University College London, London WC1E 6BT, UK 
}
\author{Marko Toroš}
\affiliation{School of Physics and Astronomy, University of Glasgow, Glasgow, G12 8QQ, UK}
\author{Antonio Pontin}
\author{Chamkaur Ghag}
\author{Sougato Bose}
\author{Tania S. Monteiro}
\author{P.F. Barker}
\affiliation{%
 Department of Physics \& Astronomy, University College London, London WC1E 6BT, UK 
}

\newcommand{\mt}[1]{{\color{blue}MT: #1}}
\newcommand{\tm}[1]{{\color{violet}TM: #1}}
\newcommand{\pb}[1]{{\color{green}PB: #1}}
\newcommand{\ji}[1]{{\color{pink}JI: #1}}

\begin{abstract}
    Motivated by the current interest in employing quantum sensors on Earth and in space to conduct searches for new physics, we provide a perspective on the suitability of large-mass levitated optomechanical systems for observing dark matter signatures. We discuss conservative approaches of recoil detection through spectral analysis of coherently scattered light, enhancements of directional effects due to cross-correlation spectral densities, and the possibility of using quantum superpositions of mesoscopic test particles to measure rare events.
\end{abstract}
\maketitle

\section{Introduction}
Despite compelling cosmological evidence for its abundance in the universe, the identification of the nature of dark matter (DM) remains an open challenge~\cite{iocco_evidence_2015,deswart_how_2017}. Whilst the subject has seen much discourse, to date, no experiment has been able to conclusively prove a dark matter event has been detected with high statistical significance. Theoretical predictions motivate a wide dark matter parameter space, and methods and strategies devised for detection depend strongly on the mass scale of the proposed candidate. 

Large scale experiments have thus far been concentrated on certain well motivated candidates (see Section~\ref{sec:Brief survey of dark matter models}), however due to limitations either on energy resolution or bandwidth, and as parameter space is ruled out in these areas, with no signal due to dark matter having been observed, the question arises: how else might we probe the fundamental structure and properties of this type of matter?

\begin{figure}[!h]
    \centering
    \includegraphics[width=\columnwidth]{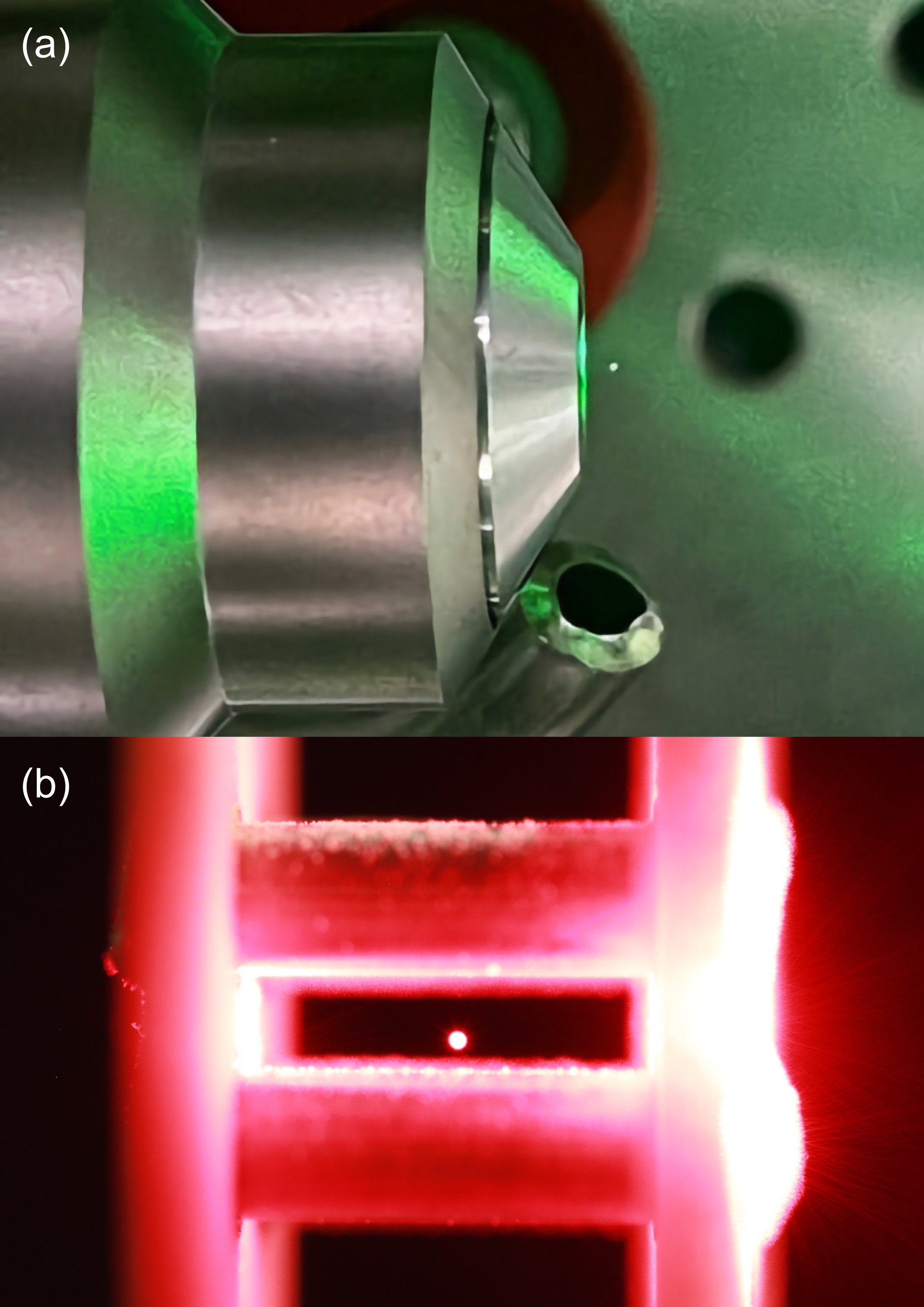}
    \caption{Levitated optomechanics: \textbf{(a)} depicts an optical tweezer setup, where a focused laser beam creates a potential well to trap and manipulate a single nanoparticle, here a silica particle with a radius of $r\,=\,70\,\text{nm}$; \textbf{(b)} shows a Paul Trap, which uses a combination of static and oscillating electric fields to confine a charged particle. The figure shows a particle with a radius $r\,=\,20\,\mu\text{m}$.}
    \label{fig:drawing_nanoparticle}
\end{figure}

A novel avenue for dark matter searches in the mid-and the low-mass sector is the use of optomechanical sensors~\cite{carney_mechanical_2021,Moore2021Searching,brady_entanglement_2023,xia_entanglement_2023,belenchia_quantum_2022,kaltenbaek_research_2023}. Levitated optomechanics in particular (illustrated in Figure~\ref{fig:drawing_nanoparticle}) promises excellent experimental control and manipulation in both translational and rotational degrees of freedom~\cite{gonzalez-ballestero_levitodynamics_2021,delic_cooling_2020,magrini_real-time_2021,tebbenjohanns_quantum_2021,pontin_simultaneous_2023}. Trapped and motion-cooled silica nanospheres~\cite{afek2022coherent,rademacher_quantum_2019}, or embedded spin systems such as diamond crystals with nitrogen-vacancy centres, can be used to perform highly precise measurements of accelerations, forces, and momentum transfers, through relative phases arising in interferometric setups~\cite{monteiro2020force,timberlake2019acceleration,lewandowski2021high,liang2023yoctonewton,winstone_direct_2018} as well as through decoherence effects. In this work, we discuss how these quantum technologies may benefit dark matter searches.

\begin{figure*}[htb]
	\begin{center}
	\includegraphics[width=\textwidth]{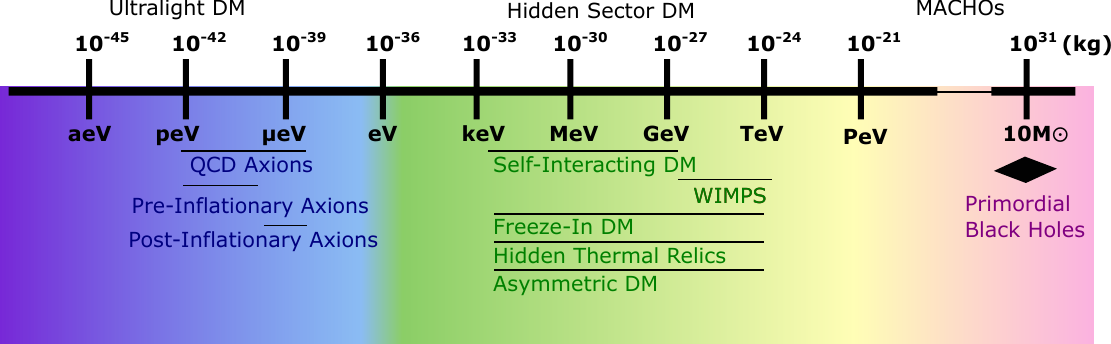}
    \vspace{-0.5cm}
	\caption{Sea of possible dark matter candidates and their representative mass scales. Ultralight Dark Matter and Hidden Sector Dark Matter are broad descriptors encompassing a range of possible theoretical candidates, a select few of which are indicated above. Ultralight candidates, shaded blue, can exhibit wave-like behaviour on cosmological scales due to their low mass, and direct detection experiments are designed to exploit this. Hidden Sector Dark Matter candidates in the green-shaded regime are characterised by interactions with standard model particles which are much weaker than the electromagnetic, strong, or weak nuclear forces. Although they form a part of the sea of candidates, pink shaded massive compact halo objects (MACHOs) are not further considered in this perspective.}
	\label{fig_dm_candidates}
	\end{center}
	\vspace{-0.5cm}
\end{figure*} 

\section{Brief survey of dark matter models}\label{sec:Brief survey of dark matter models}

The anticipated sensitivity of a given experimental setup to any particular dark matter candidate is strongly model-dependent. A potential dark matter signal is governed by a few crucial parameters. The candidate's mass sets the kinematics of interaction, and hence local number density and relative velocity with respect to Earth of its theorised particles in the galactic halo~\cite{read_local_2014,fox_interpreting_2011,kavanagh_measuring_2021}. Additionally, any predicted non-gravitational couplings will further affect interactions with standard model particles~\cite{hochberg_new_2022}.

In the standard Lambda cold dark matter model ($\Lambda\text{CDM}$), formulated on the assumption that the universe is, in an approximate sense, flat on large scales, isotropic and homogeneous, dark matter is predicted to make up $\sim\,26.8\,\%$ of the mass-energy content of the universe. This dark matter is non-relativistic, non-baryonic, weakly interacting, and expected to consist of massive particles. A formerly strong contender surfacing naturally through supergravity and supersymmetric models are weakly interacting massive particles (WIMPS), assumed to interact with Standard Model particles through the gravitational and electroweak force. However, in the predicted mass range of 1~GeV-100~TeV, no such particles have been detected so far, and many WIMP models have been excluded by experiments~\cite{aalbers_first_2023}. 

Whilst the standard cosmological model has been successful in explaining a variety of astrophysical observations, it also suffers shortcomings, a prominent one being the statistically significant disagreement between theoretically derived and experimentally measured values of the Hubble constant. Among many such deviations~\cite{del_popolo_small_2017}, discrepancies like the aforementioned open the field of possible dark matter models.

An alternative model that mimics the behaviour of CDM at larger scales but exhibits vastly different phenomenology at small mass scales, supposes the existence of ultralight bosonic dark matter (ULDM) with wave-like behaviour. Oscillations of such dark matter waves give rise to time-dependent variations of the electron to proton mass ratio, and fundamental constants such as the fine structure constant, both of which are readily measured with atomic clocks. Additionally a lower limit on the mass of fermionic dark matter $m_{\text{DM}, \text{f}}\gtrsim\,100\,\text{eV}$ is imposed by the Tremaine-Gunn bound~\cite{tremaine_dynamical_1979}.

Motivated by the expected experimental signal and for ease of discussion, we will split the particulate candidates broadly into two categories. If the number density is sufficiently high (occurring for bosonic dark matter masses $m_{\chi}\,\lesssim\,1\,\text{eV}$) then the field can be approximately described by a Bose–Einstein condensate, exhibiting wave-like behaviour and coherence over a timescale set by the natural frequency of the field. For candidates with masses larger than $\sim\,1\,\text{eV}$ however, the interactions will be impulse-like in nature, with a momentum transfer dependent on the mass of the candidate.


\subsection{Ultralight candidates}

Introducing lighter mass dark matter candidates into the standard model can often be achieved through the addition of a single ultralight field, commonly in the form of a spin 0 scalar or pseudo-scalar field, or a spin 1 vector or axial-vector field~\cite{bernal_scalar_2006}. The choice of these specific spin values is informed by theoretical models and observational constraints. Higher spin values, such as spin 2 and above, introduce significant complexities in dark matter models, often leading to theoretical challenges and inconsistencies, particularly in their coupling and interaction with gravitational fields~\cite{criado_dark_2020, asorey_higher_2010}. In reference to Figure~\ref{fig_dm_candidates}, these candidates typically lie in the $\lesssim\,1\,\text{eV}$ mass range. The interaction mechanism of these fields with standard model particles dictates the experimental observables, which may include variations in fundamental constants, spin torque coupling, electromagnetic effects, or dark matter-induced accelerations.

These fields are often treated as having monochromatic oscillations due to their coherent wave-like nature within a narrow frequency range, arising from their extremely light mass~\cite{ferreira_ultralight_2021}. The term `monochromatic' in this context refers to this narrow energy spread. Despite the diversity of the possible experimental observables, certain induced matter effects, such as differential accelerations in materials with different nuclear compositions, are detectable with levitated sensors. Both scalar and vector fields can couple to standard model nucleons, making levitated mechanical sensors of varying material composition well-suited for narrow-band searches for these dark matter candidates (for an expanded discussion, see Section~\ref{sec:Levitated optomechanics}).

\subsection{Hidden Sector candidates}

For a given local mass density ($\rho\,\sim\,0.3\,\text{GeV/cm}^3$), heavier dark matter candidates (in the $>1\,\text{eV}$ mass range as shown in Figure~\ref{fig_dm_candidates}) have a lower number density, ruling out wave-like behaviour due to insufficient density. Their interactions with standard model particles are impulse-like in nature, leading to energy deposition and recoil of the target nucleus. Typically, dark matter interactions are inferred from this energy deposition, observable through atomic excitation or scintillation. However, two issues arise for such candidates: first, the momentum transfer $p$ during the interaction is small ($p\,=\,m_{\chi}\,v_{\chi}$, with $v_{\chi}$ representing the velocity of dark matter candidates, assumed to be non-relativistic), leading to a requirement for highly sensitive detection thresholds. Second, the rarity of these recoils\,—attributed to the low local number density and theorised small interaction cross-sections—\,necessitates large target sizes for feasible experimental timeframes. To address these challenges, large-scale noble element detectors are employed to exploit the advantage of size~\cite{aalbers_first_2023}, whilst small mechanical sensors are utilised for their low detection thresholds~\cite{hochberg_new_2022}.

\section{Levitated optomechanics}\label{sec:Levitated optomechanics}
The roots of levitated optomechanics trace back to the pioneering work of Arthur Ashkin in 1971 with the development of optical tweezers and their application to the levitation of a dielectric particle in vacuum~\cite{ashkin1971optical}. In recent years, the field has rapidly matured and the quantum ground state of the centre-of-mass motion has been reached in one and two dimensions~\cite{delic_cooling_2020,tebbenjohanns_quantum_2021,magrini_real-time_2021,piotrowski2023simultaneous}. 

Levitated optomechanics~\cite{millen_optomechanics_2020,gonzalez-ballestero_levitodynamics_2021} is a very appealing field to employ for dark matter searches. Experiments are typically performed in ultrahigh vacuum (with gas pressures as low as $10^{-11}\,\text{mbar}$)~\cite{dania2023ultrahigh}, thus suppressing the Brownian motion noise from collisions with ambient gas. Levitated systems interact predominantly with the optical, electric or magnetic fields employed to trap, control and measure their motion, as well as any residual gas molecules. As they are largely isolated from the environmental sources of noise, they offer unparalleled force sensitivities, down to zeptoNewton scales and below~\cite{liang2023yoctonewton}. 

Another notable advantage of the levitated systems is that the motional modes ($x,\,y\,\text{and}\,z$) for the 3D centre of mass motion lie along a Cartesian axial frame, allowing simultaneous probing of all the components of an applied external force, and determination of its precise orientation. It should be emphasised that although the small size of levitated sensors relative to standard noble element detectors may imply reduced coupling between dark matter and the oscillators, this may be compensated for, especially by the ability for directional detection and, depending on the candidate, additional enhancement factors in the cross-section due to coherent scattering.

{\bf Optical levitation} has been the most explored technique thus far. Here, silica nanospheres of diameter $70\,-\,200\,\text{nm}$ have been trapped with a mass of $\approx\,10^{-18}\,-\,10^{-16}\,\text{kg}$ and oscillation frequencies on the order of $30\,-\,300\,\text{kHz}$.
These dielectric nanoparticles are levitated in optical tweezer traps obtained via tightly focusing laser beams: typically, a Gaussian laser beam is tightly focused by a high numerical aperture (NA) lens to form a diffraction-limited beam focus. The nanoparticles are then attracted to the beam focus by a gradient force, which ultimately traps them.

{\bf Free-space vs. cavity optomechanics:} two distinct approaches are widely used within optical levitation for control: (i) with active control, time-dependent "feedback" forces (usually electric) are applied to damp the motion along the $x,\,y\,\text{and}\,z$ axes allowing quantum cooling regimes to be achieved in 1D "free-space"~\cite{tebbenjohanns_quantum_2021,magrini_real-time_2021}, (ii) with passive control, the tweezer trap is coupled to an optical cavity comprising of a pair of highly reflective mirrors. The resultant coupling of the motions to the optical mode of the cavity has provided a powerful method for cooling and read-out, allowing quantum ground state cooling regimes both in 1D~\cite{delic_cooling_2020} and 2D~\cite{piotrowski2023simultaneous}. 
 
In the latter case, the motion is detected via the cavity mode light that is transmitted through the cavity, then interfered with a strong reference beam via a technique termed homodyne or heterodyne detection. 

In the former case, the $x,\,y,\,z$ motions are detected in real-time interferometrically by measuring the interference between the light scattered from the particle and the light used to illuminate it. This causes modulations of the propagating light which are detected using photodiodes, converting this to a voltage modulation. Calibration of this voltage signal gives a real-time measurement of the particle's position~\cite{Hebestreit2018Calibration}. For non-spherical particles, their rotational motion modifies the polarisation of the scattered light, which can also be measured~\cite{nieminen2001optical}. 

In both cases, Fourier transformation of the time signal to the frequency domain provides information on the particle's oscillation frequency, temperature and damping. Once a steady state is achieved, the power spectral density (PSD) is a standard tool of analysis. 

{\bf The minimum detectable force} in levitated optomechanics is comparable to the background thermal noises (such as the Brownian noise from background gas) that heat the particle's motion. A widely used expression for the minimum force is readily obtained~\cite{gosling2023sensing} \mbox{$F_\text{min}\,=\,\sqrt{S_\text{th}\,b}\,=\,\sqrt{2\,k_B\,m\,\Gamma\,T_{CM}\,b}$}, where $m$ is the mass of the oscillator, $k_B$ is Boltzmann's constant, $\Gamma$ is the damping rate determined by the active or passive damping method employed, and $T_{CM}$ is the centre-of-mass temperature. The parameter $b$ is the measurement bandwidth, which in levitated experiments is effectively the inverse of the experimental integration time.
A simple reading of this expression reveals the surprising fact that quantum cooling is unimportant for force sensitivity: the product $\Gamma\,T_{CM}$ is constant since $T_{CM}\,\propto\,\Gamma^{-1}$. Hence some of the highest force sensitivities to date have been demonstrated near $T_{CM}\,\sim\,300\,\text{K}$ i.e. with modest cooling, used simply to stabilise levitation~\cite{liang2023yoctonewton}.

Nevertheless, a goal for many experiments is to reach quantum regimes where quantum coherent effects, possibly in the multi-mode and multi-particle regimes, offer further enhancements.
In addition, cooling remains important for sensing using correlations between motional modes ~\cite{gosling2023sensing}.

{\bf The standard quantum limit (SQL)} of displacement sensing is a key goal for non-levitated and levitated optomechanics. For the harmonic oscillator modes of the mechanical motion, their zero point fluctuation $x_{zpf}\,=\,\sqrt{\hbar/(2\,m\,\omega_x)}$ sets a fundamental limit to displacement sensing. However, any realistic sensing must, in addition, allow for the measurement disturbance. For optical sensing, employing a weaker optical probe enhances the effect of quantum shot noise due to the random arrival of the photons; conversely, using a stronger optical probe reduces the effect of shot noise but enhances the back action noise (from photon recoil in free-space experiments or dynamical coupling to the cavity mode for cavity-based optomechanics). Minimising the combined effect of shot noise and back action noise sets the so-called standard quantum limit for displacement sensing. To date, measurements within about $20\,\%$ of the SQL have been achieved in a levitated setup~\cite{magrini_real-time_2021}.

It is nevertheless still possible to go beyond the SQL by employing squeezed light modes arising from the coupled cavity-mechanical motion dynamics (ponderomotive squeezing)~\cite{Mason2019Continuous}; or by directly using a squeezed source of light as a probe~\cite{Gonzalez-Ballestero2023Suppressing}. As forces are measured via the displacements, these advances are relevant to force sensitivity. 


\section{Levitated Sensors For Dark Matter}\label{sec:Levitated Sensors For Dark Matter}

\begin{figure*}[t]
    \centering
    \includegraphics[width=.8\textwidth]{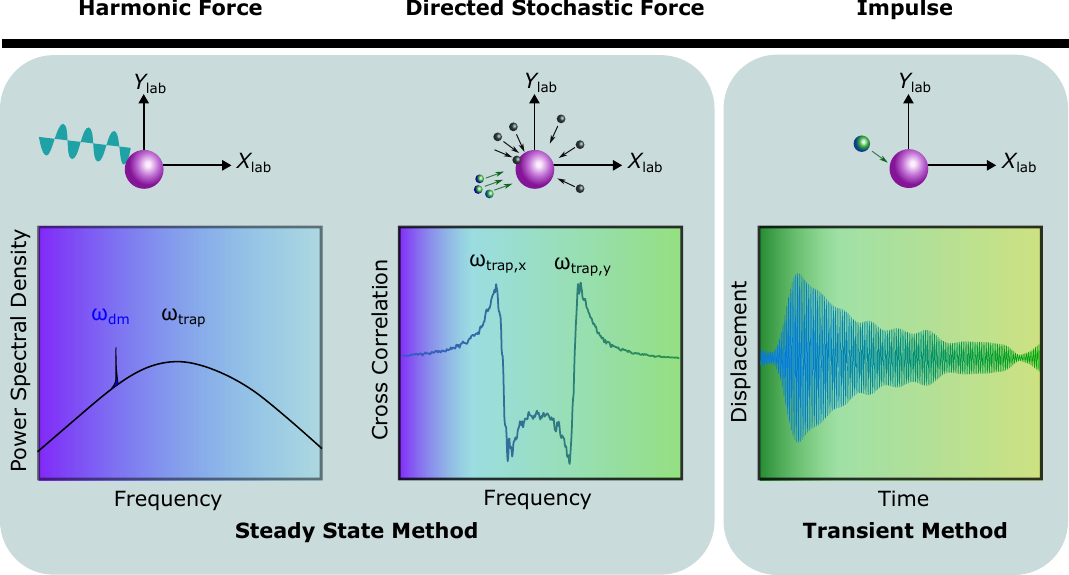}
    \caption{Typical forms of the measured signals due to the different types of forces detected by levitated sensors. How to detect the force is dependent upon the time of interaction and the type of force. For harmonic forces, the PSD of the oscillatory mode is modified to include an additional peak at the frequency of the harmonic drive. The height of the peak is dependent upon the size of the force. For a directional stochastic force, the cross-correlation mechanical spectra yield additional information on the orientation quadrant of the force and are an unambiguous signal of a directional force. For impulse-like events, the interaction times are shorter than the oscillation period of the particle and are hence investigated in the time domain. The impulse results in a sharp displacement of the particle's position which then decays at a rate dependent upon the damping experienced by the particle.}
    \label{fig:forces}
\end{figure*}

\textbf{The types of forces} typically detected via levitated sensors are constant, static forces; impulse, "kick"-like forces; and harmonic forces, oscillating at a particular frequency. A more recent study also investigates the measured effect of a directed, stochastic force~\cite{gosling2023sensing}. How these forces are measured depends upon the type of force and the length of interaction. A pictorial representation of the particle interaction for each force, and the experimental signals to be measured as a result, can be seen in Figure~\ref{fig:forces}.

A constant force results in a change in the probability distribution of the position of the particle, i.e. a shift in its mean position.

For an impulse-like event, the particle experiences a transfer of momentum over a time much shorter than its oscillation period, resulting in a sudden increase in amplitude of its oscillations. These short, transient interactions are hence considered temporally.

When the interaction time is longer than the oscillation period it is instead best to study the dynamics in the frequency domain, through changes in spectral densities such as the PSD, where random fluctuations due to residual gas and photon scattering are averaged over. In the presence of a harmonic force, a sharp "delta" peak appears in the PSD at the frequency of the harmonic oscillations, in addition to the (usually broader) peak at the trap frequency. The height of the sharp peak is dependent on the size of the force. For a stochastic broadband force, the force operates over a wide range of frequencies and results in an increase in the height of the PSD, with the increase dependent on the size of the force. However, when this stochastic force is directional, additional information can be extracted by considering the correlations induced via the cross-correlation mechanical spectra. This is discussed further in Section~\ref{sec:Levitated Sensors For Dark Matter}.

\textbf{Decoupling from the environment} enables levitated oscillators to exhibit high Q factors, making them excellent force sensors, with a force sensitivity of $4.34\,\text{zN/Hz}^{1/2}$ having been measured experimentally with an optically levitated nanoparticle~\cite{liang2023yoctonewton}.

Whilst optical levitation through the use of a high-NA lens has achieved the highest force sensitivity, the size and material used for the test particle are limited. The size of the particle in a high-NA optical tweezer is limited by the increased scattering force with an increased size to the point that the particle can no longer be stably trapped. The possible materials that can be used are limited due to absorption of the high-intensity light used to levitate the particle. This absorption results in the increase in internal temperature at high vacuum as the gas is no longer present in the chamber to thermalise the levitated particle. This can result in internal temperatures up to 1000\,K \cite{hebestreit2018measuring} and lead to the melting of the nanoparticle \cite{rahman2016burning}.

However, the size range of possible particles that can be levitated is expanded by utilising gravito-optical traps~\cite{moore2014search,arita2022all,kawasaki2020high}, or trapping via electric~\cite{bullier2021quadratic,alda2016trapping,dania2022position}, or magnetic fields~\cite{Brown2023Superfluid,vinante2020ultralow,gieseler2020single,latorre2023superconducting}. These techniques allow the levitation of micron-sized spheres~\cite{monteiro2020force,timberlake2019acceleration,lewandowski2021high}, and this large \mbox{mass $\sim\,10^{-6}\,-\,10^{-9}\,\text{g}$} has enabled a measured acceleration sensitivity of $\num{9d-12}\,\text{g/Hz}^{1/2}$~\cite{hofer_highq_2023}. The only requirement for electric field levitation is a specific charge-to-mass ratio determined by the trap design. Hence there is significantly more freedom in the choice of material for the levitated object, allowing for higher densities and use of materials with unique properties, for example, NV centers~\cite{delord2020spin}. For electrical and magnetic trapping, the oscillation frequencies can be as low as a few Hz up to approximately tens of kHz. 

The capacity to confine particles with substantially concentrated mass, thereby enhancing the coupling potential, is especially beneficial for bosonic ULDM searches~\cite{carney2021ultralight}. Diamagnetically levitated micromechanical oscillators have been proposed to investigate ULDM, utilising the coupling to the $B-L$ charge (baryon number minus lepton number)~\cite{li2023search}. This study predicts improved limits to the $g_{B-L}$ coupling constant by an order of magnitude compared to previous results, and sensitivity to $g_{B-L}\,\sim\,\mathcal{O}(10^{-25})$ in the frequency range of $0.1\,\text{to}\,100\,\text{Hz}$ (see Table~\ref{tab:DMtable}).

A levitated particle benefits from a high mass concentration which couples well to the dark matter candidate whilst being small enough that interactions with a light candidate can be considered coherent across the particle. 

Additionally, whilst a single mechanical oscillator has maximum sensitivity at a single resonant frequency, an array of levitated particles, facilitated through the use of electrodynamic traps~\cite{bykov_3d_2023,penny2023sympathetic}, allows for probing of a much wider ultralight dark matter spectral range. This would allow for a broadband search for dark matter of this type, increasing interaction cross-section and reducing the time needed to conduct a search of this type.

\textbf{Non-spherical, asymmetric particles} have also been investigated. Through levitation, this kind of particle is free to rotate around its principle axes and can be driven by circularly polarised light to rotate at $\sim\,\text{GHz}$ frequencies~\cite{ahn2018optically,reimann2018ghz,zielińska2023spinning}. These ultrahigh rotation frequencies make these devices excellent torque sensors~\cite{ahn2020ultrasensitive} with a torque sensitivity of $(4.2\,\pm\,1.2)\,\times\,10^{-27}\,\text{Nm/Hz}^{1/2}$ having been achieved. Combining all the different systems mentioned before means that the levitated optomechanical toolbox is able to investigate a wide range of phenomena with high sensitivity.

Unlike atomic systems, macroscopic particles also possess an internal bulk temperature. This internal temperature can be measured via changes in the particle's refractive index~\cite{hebestreit2018measuring}, changes in fluorescent light~\cite{Rahman2017Laser,laplane2023inert,Zhang2023Determining}, or microwave resonances~\cite{Rivière2022Thermometry}. This internal temperature can also be cooled by solid-state laser refrigeration to as low as $130\,\text{K}$~\cite{Rahman2017Laser,laplane2023inert} with further enhancements proposed to reach cryogenic temperatures~\cite{Gragossian2016Astigmatic}.

\textbf{A variety of searches for different dark matter candidates} have already been undertaken using levitated optomechanics. In general, interest in lower-mass candidates has grown as parameter space for direct detection of weakly interacting massive particles (WIMPs) is further excluded by large-scale noble element detectors~\cite{aalbers_first_2023}.
Levitated optomechanical systems possess concentrations of mass ranging approximately from $10^{10}\,-\,10^{15}$ atomic masses, significantly enhancing the probability of interaction with select dark matter candidates.

The first realisation of an experimental search for dark matter with levitated particles utilised not an interaction with galactic dark matter candidates, but millicharged dark matter bound to Standard Model particles~\cite{afek2021limits}. Control of the net charge of a levitated object can be obtained through the use of UV light or corona discharge~\cite{ricci2019accurate}. The neutralisation of a microsphere has enabled the possibility of excluding partial charges larger than $10^{-4}\,e\,\text{(GeV\,to\,TeV)}$, marking a two-order-of-magnitude enhancement compared to the state of the art of the time. 
~\cite{monteiro2020search}. This specific model benefits from the microsphere being a highly localised collection of nucleons enhancing the interaction cross-section and experimentally excluding new parameter space, with $\sigma_{\chi\,n}\,\text{[cm}^2\text{]}\,\sim\,10^{-28}$ for $10^3$\,--\,$10^4\,\text{GeV}$ dark matter mass. Impulse sensitivity can be increased by considering levitated nanoparticles at the standard quantum limit, when $\Delta\,p\,=\,\sqrt{\hbar\,m\,\omega_0/2}$ where $m$ is the mass of the oscillation and $\omega_0$ the oscillation frequency~\cite{clerk2004Quantum}. As mentioned, working with levitated nanoparticles also offers enhancement via coherent scattering across the size of the nanoparticle~\cite{afek2022coherent}, enabling the investigation into low mass particle dark matter with projected sensitivity of $\sigma_{SI}\,\text{[cm}^2\text{]}\,\sim\,10^{-30}\,(m_{\chi}\,\sim\,0.1\,-\,100\,\text{MeV)}$ for a 10 by 10 array. Given this increase in cross-section, levitated nanoparticles have also been proposed, in conjunction with the use of scintillator panels instrumented with Silicon Photomultiplier (SiPM) arrays, for investigating heavy sterile neutrinos (another dark matter candidate), through examination of the products of radioactive decays and reconstructing neutrino mass~\cite{carney2023searches}. This method has a projected sensitivity of \mbox{$\left |U_{e4} \right |^2\,\sim\,10^{-4}\,-\,10^{-6}\,$}$\text{(for}\,m_{\chi}\,\sim\,10\,\text{keV\,to}\,1\,\text{MeV}$) for 1000 nanospheres. Both approaches utilise the advantages which come with the levitation of multiple particles. Levitated sensors are becoming increasingly sensitive to impulse kicks to such an extent that they are approaching the ability to resolve individual gas collisions~\cite{magrini_real-time_2021,barker2023collisionresolved}.

Levitated ferromagnets in a superconducting trap have been shown to be excellent magnetometers, surpassing the energy resolution limit and demonstrating the highest reported sensitivity in terms of bare energy per unit volume~\cite{ahrens2024levitated}. This level of sensitivity to magnetic fields has been suggested for the study of axions and axion-like dark matter via its coupling to the electron spins of the ferromagnetic magnetometer with a projected sensitivity of \mbox{$g_{aee}\,\sim\,\mathcal{O}(10^{-14})\,(m_{\chi}\,\sim\,10^{-13}\,\text{to}\,10^{-18}\text{eV)}$}. Levitated superconductors in a magnetic trap have also been proposed for investigating axions and dark photons via perturbations of the levitated superconductor's equilibrium position due to the oscillating magnetic field caused by the ultralight dark matter~\cite{higgins2023maglev}. For axions, the projected sensitivity is \mbox{$g_{a\gamma}\,\text{[GeV}^{-1}]\,\sim\,\mathcal{O}(10^{-10})\,(m_a\,\sim\,10^{-12}\,\text{to}\,10^{-14}\text{[eV])}$} and the projection for dark photons \mbox{$\epsilon\,\sim\,\mathcal{O}(10^{-8})\,(m_{A'}\,\sim\,10^{-12}\,\text{to}\,10^{-14}\,\text{[eV])}$}. Magnetically levitated objects have also been proposed and utilised to investigate dark energy, ruling out the basic chameleon model as a candidate for dark energy~\cite{Betz2022Searching,Yin2022Experiments}.

\section{Multi-mode levitated optomechanics}\label{sec:Multi-mode levitated optomechanics}
\begin{figure}
    \centering
    \includegraphics[width=\columnwidth]{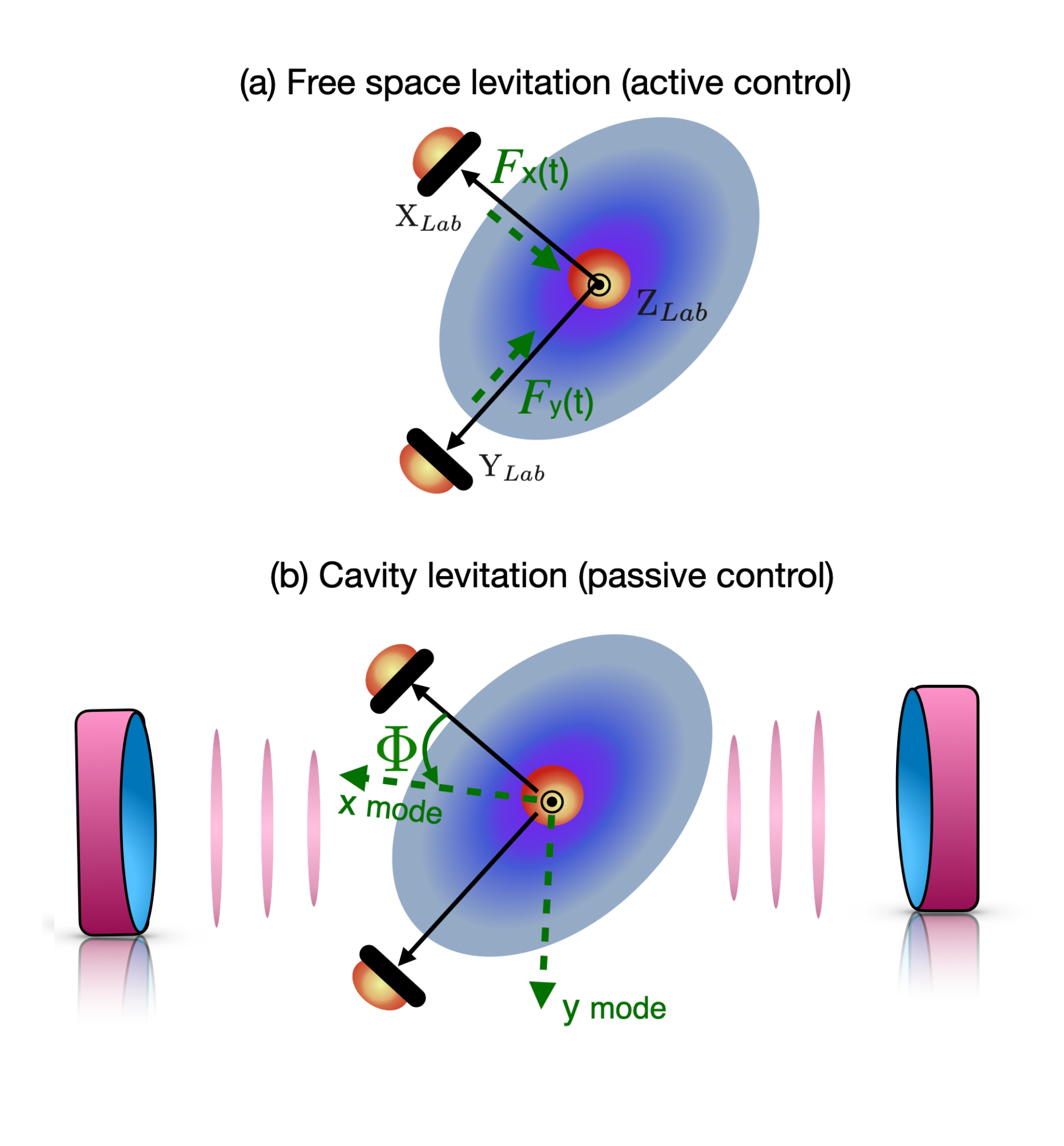}
    \caption{Two major setups in optical levitation. \textbf{(a)} The nanoparticle is levitated at the focus of an optical tweezer. The $x,\,y,\,z$ axes are set by the tweezer potential (blue). The motion is detected from the scattered light and used to control feedback forces along $x,\,y$ and $z$ that cool the motion. To date, only quantum cooling in 1D along the axial direction ($z$) has been achieved with this setup, but sensing of forces down to zeptoNewtons (and even yoctoNewtons) has been demonstrated. \textbf{(b)} In this technique, the tweezer trap is surrounded by an optical cavity. The cavity optical mode can cool and read out the motion very effectively; but cavity back-action hybridises the $x,\,y,\,z$ modes. Hybridisation here implies a geometrical rotation of the normal modes and (as shown here) the $y$ mode becomes a dark mode (uncoupled from the cavity field), and the $x$ is a bright mode. 2D quantum cooling of the $x,\,y$ plane is possible only in a “Goldilocks” zone where the coupling of modes to the cavity is strong enough for cooling, but not strong enough for dark modes. Directional force detection requires control of the mode rotation angle so modes remain aligned with the lab detectors.}
    \label{fig:DM1}
\end{figure}

An extremely useful feature of levitated probes for dark matter searches is that the measured signals have a straightforward geometrical interpretation. For a spherical particle, the typical signals measured correspond to the particle's motional normal modes in a Cartesian frame ($x,\,y,\,z$). This simple description is useful when considering dark matter interactions. A dark matter interaction should exhibit a known directionality deriving from the Earth's motion relative to the galactic reference frame~\cite{ahlen2010case}. This directionality enables discrimination against background events that either a) interact in a direction not expected or b) exhibit no directionality at all.

Levitated optomechanics is thus uniquely suited to searches looking to measure $x,\,y,\,z$ components of an external force. However, it is essential that the $x,\,y,\,z$ motional modes are accurately aligned with the frame $X_{lab},\, Y_{Lab},\,Z_{Lab}$ of the laboratory detectors. 

For a free-space scheme, where the particle is held in a tweezer trap, 
the tweezer axes must be carefully aligned with the detectors as illustrated in Figure~\ref{fig:DM1}. Any misalignment will result in a spurious (uninteresting) correlation between the normal modes that masks more interesting, externally induced correlations between the modes.

For cavity-based schemes, the strong coupling between the cavity mode and the motion, as well as the sensitivity of cavity-based detection, is extremely advantageous for quantum cooling: the first quantum ground state experiments were reported using a cavity set-up~\cite{delic_cooling_2020} in 1D, corresponding to the cavity axis.
However, the alignment between the orientation of the mechanical normal modes and the laboratory detection axes is complicated by the back-action from the cavity optical mode. A study of the {\bf cavity-mediated coupling} of the mechanical modes showed that cross heating due to mode hybridisation could significantly affect 1D quantum cooling experiments~\cite{toros_quantum_2020}.

\begin{figure*}
    \centering
    \includegraphics[width=\textwidth]{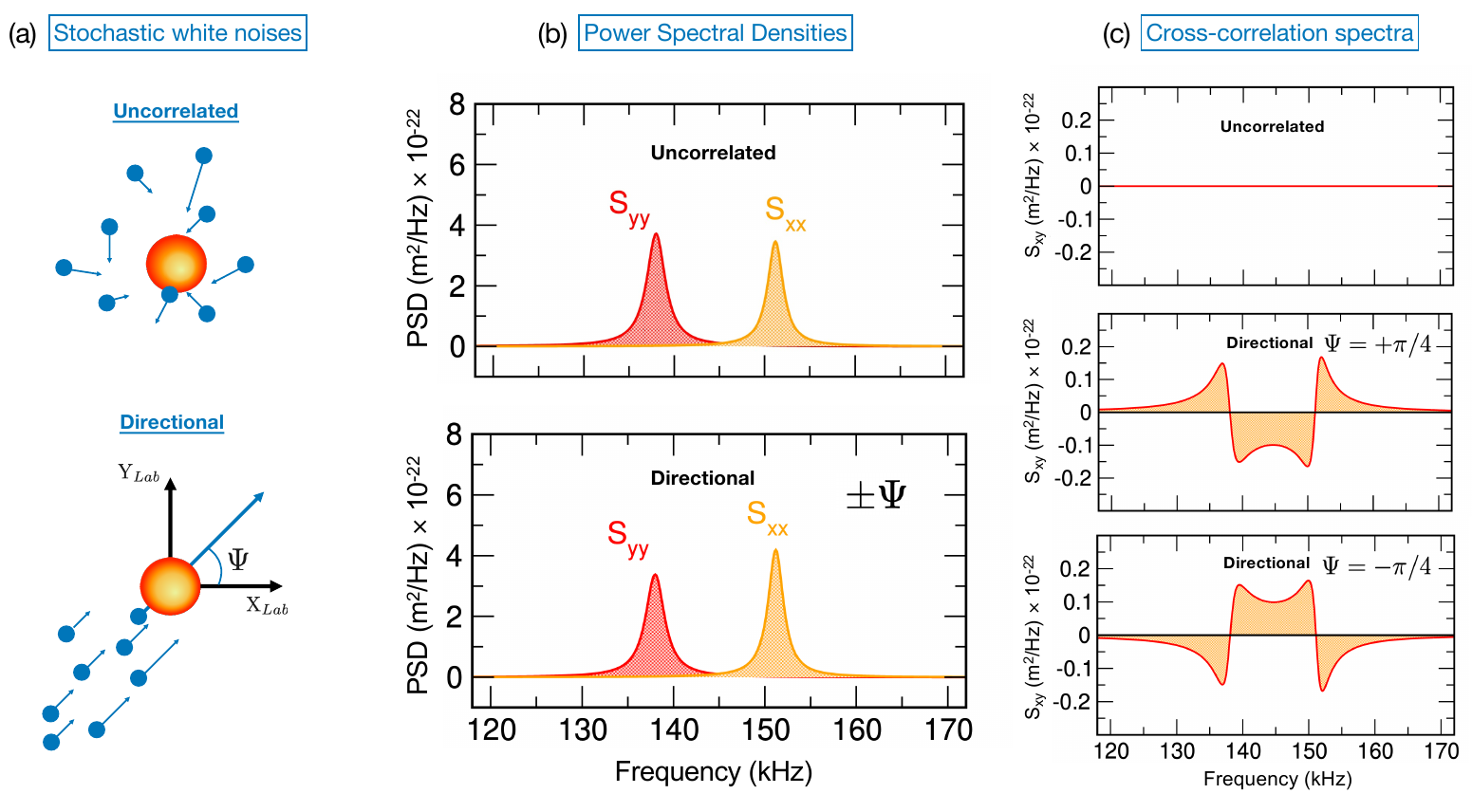}
    \caption{Detection of directional noise sources with a levitated system. \textbf{(a)} Schematic illustration of an uncorrelated noise source, with a white frequency spectrum (upper diagram) as well as a stochastic, uncorrelated stream of noise with a definite orientation angle $\Psi$ in the $x-y$ plane (lower diagram). \textbf{(b)} Shows the corresponding power spectral densities (PSDs) for $x$ and $y$. The effect of uncorrelated noise (upper plot) is analogous to thermal noise baths acting on the particle: it acts to heat the $x$ and $y$ modes of the motion. The directed noise (lower plot) may result in different rates of $x,\,y$ heating, but is also easily masked as a simple change in the thermal heating rates. It is insensitive to quadrant (the PSDs for $\pm\,\Psi$ are both plotted, and are equal). \textbf{(c)} In contrast, the cross-correlation spectra give an unambiguous signature of the presence of a directional noise source. The cross-correlation spectra are absent if the noise is uncorrelated (upper plot), but appear for a directional noise source (middle and lower plots), and give a cross-correlation spectrum that is sensitive to orientation quadrant (as the spectra for $\pm\,\Psi$ are not equivalent).}
    \label{fig:CrossCorr}
\end{figure*}

Hybridisation between the mechanical modes results in a geometric rotation of the normal mode~\cite{toros_coherent_2021} orientation relative to the lab frame. If both modes are well coupled to the cavity, one mode rotates so it becomes a {\bf dark mode}, orthogonal to the cavity field, so cannot be cooled. The other mode, in contrast, rotates into alignment with the cavity, forming a {\bf bright mode} that is strongly cooled. This is
illustrated in Figure~\ref{fig:DM1}. A study of 2D quantum cooling of the $x-y$ plane~\cite{toros_coherent_2021} found it was only possible in a {\bf ``Goldilocks'' zone}, where cavity coupling is strong enough for quantum ground state cooling, but not strong enough for the formation of dark/bright modes. A recent experiment~\cite{piotrowski2023simultaneous} in the Goldilocks parameter regime achieved 2D quantum cooling.

Another recent experimental study investigated the back-action induced rotation of the particle's normal modes~\cite{pontin_controlling_2023}. A particular trapping position where the back-action induced rotation is cancelled by an interference with the trapping field~\cite{toros_quantum_2020} was demonstrated. At this point, externally induced correlations are exposed since the misalignment caused by these (parameter-sensitive) mode rotations is eliminated. Levitated nanospheres in a Fabry-Perot cavity have been proposed to investigate charged dark matter bound to ordinary matter via the cavity's output light~\cite{asjad2023charged}.

\subsection{Directional sensing via cross-correlations between modes}

In a recent study~\cite{gosling2023sensing}, it was proposed and experimentally verified that there is a technique that is well adapted to detect -- in the steady state -- a dark matter candidate that appears as a source of 
{\bf stochastic} white noise, but from 
a preferred direction. Such a source might normally be difficult to distinguish from the ordinary thermal noises that heat the particle, especially if the source is weak.

Figure~\ref{fig:CrossCorr} illustrates this concept. The standard approach to detect the effect of thermal noises uses PSDs $S_{xx}(\omega)\,=\,\langle |x(\omega)|^2\rangle\,\text{and}\,S_{yy}(\omega)\,=\,\langle |y(\omega)|^2\rangle$ where the angled brackets denote an average over an ensemble of measured spectra. The area under the curve yields the
temperature of the $x$ and $y$ modes respectively. The behaviour of the PSDs is qualitatively similar whether the bath is uncorrelated or not: the directional force appears simply as yet another thermal source. There is a quantitative 
effect but it may easily be confused with a change in multiple other thermal heating sources present in the experiment. A directional signal can also be easily masked by uncertainties in directional heating rates that arise from slight particle asymmetries.

However, the signature of the {\bf cross-correlation spectrum} $S_{xy}(\omega)\,=\,\langle\,x^*(\omega)\,y(\omega)\,\rangle$ is unambiguous. As shown in Figure~\ref{fig:CrossCorr}(c), provided there are no misalignments between the laboratory detectors and the normal modes, an uncorrelated stochastic noise bath yields a zero cross-correlation spectrum. However, if the noise has a preferred direction, there is a distinctive spectral signature. Since $S_{xy}(\omega)\,\propto\,\cos\Psi\,\sin\Psi\,\propto\,\sin2\Psi$, the spectra are sensitive to quadrant: their shapes flip for $\pm\,\Psi$.
 
Cooling is important for detection of cross-correlation spectra: in order to generate $x-y$ cross correlations, one requires that the damping rates $\Gamma_{x,y}\,\sim\,|\omega_x-\omega_y|$~\cite{gosling2023sensing}. This is in sharp contrast to standard force sensing where
the sensitivity is largely independent of the experimental cooling rate (see discussion of $F_{min}$ in Sec. III).

Additionally, a directional signal due to dark matter will modulate yearly due to the Earth's movement through the galactic reference frame, resulting in a sign change in the correlation signal when the direction passes through $\Psi\,=\,0$.
This sign change (due to the changing direction of a directional stochastic noise source) has already been measured experimentally~\cite{gosling2023sensing}.
Hence, the correlations between different motional modes are potentially valuable as a discriminant in searching for directional dark matter signals.

In summary, the cross-correlation technique complements current methods. It offers a promising new approach to detect a dark matter signal that is not a simple harmonic force or a source of strong but infrequent recoils, but rather a source of white noise that may be missed with current methods. It might correspond to a stream of weak impacts, individually undetectable but frequent enough to generate correlations. Alternatively, it might correspond to a broadband spectrum of harmonic coherent sources, across a range of frequencies.

\section{Mesoscopic Matter-Wave Interferometers}
A recently popularised approach for direct searches for dark matter candidates, extending beyond the current experimental status of levitated optomechanics, is rooted in the creation of \textbf{quantum superposition states}. Novel experimental protocols using embedded single spins~\cite{bose_spin_2017}, the typical example of one such physical system being nitrogen-vacancy centers in diamond crystals, may be utilised to induce a separation of the motional wavepacket of nanocrystals. Similarly, optical field couplings in trapped atom-nanoparticle hybrid systems~\cite{toros_creating_2021}, or tunnelling effects in ferrimagnetic nanoparticles can be explored to produce non-classical states~\cite{rahman_large_2019}.

\begin{figure}
    \centering
    \includegraphics[width=.5\textwidth]{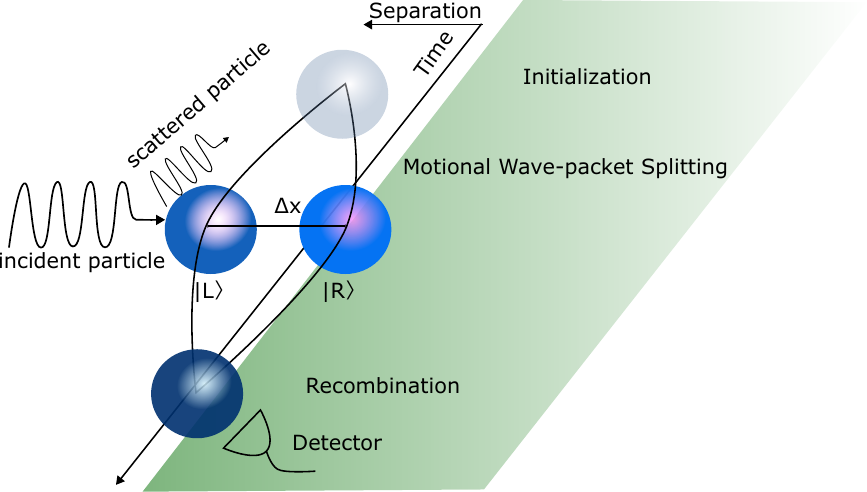}
    \caption{Illustration of a mesoscopic matter-wave Stern-Gerlach-type interferometry setup for particulate matter detection. An incoming particle scatters from a superposed quantum object, such as a nano-crystal in a superposition of center-of-mass motional states $|R\rangle$ and $|L\rangle$, resulting in a transfer of momentum to the quantum object. Upon recombination of the wave-packet and completion of the interferometry, the scattered particle can be sensed through the relative phase accumulated between the components $|R\rangle$ and $|L\rangle$, as well as through a decoherence contribution manifesting in a reduction of interferometric fringe visibility. The former effect may dominate over the latter, depending on the characteristics of the particulate environment and the experimental parameters such as the superposition size $\Delta\text{x}$.}
    \label{fig:SG-Interferometer}
\end{figure}

Regardless of the choice of the physical system, the methodology for future matter-wave interferometric experiments is largely the same. Once the mesoscopic quantum system is initialised in a superposition of its centre-of-mass motion and a state of the form \mbox{$|\psi(t)\rangle=\frac{1}{\sqrt{2}}\bigl(|\psi_{L}(t)\rangle+|\psi_{R}(t)\rangle\bigr)$} has been created, with $L$ and $R$ symbolically representing left- and right-propagating wavepackets, it is often beneficial to keep the separation $\Delta\,\text{x}$ between the superposed components constant during most of the experiment time. Relative phases $\Delta\phi$ accumulated between the arms of the superposition can finally be measured after the interferometer is closed. In quantum systems featuring embedded spins, the phases are particularly easily inferred from measurements performed solely on the spin states.

Stern-Gerlach interferometers based on spin-mechanical mesoscopic objects have been richly discussed in the context of general relativity and gravitational physics, where they have been proposed as sensors for gravitational waves, spacetime curvature~\cite{marshman_mesoscopic_2020}, tests of the equivalence principle~\cite{bose_entanglement_2023}, and quantum gravity theories~\cite{bose_spin_2017,marletto_gravitationally_2017}. In addition to their suitability as detectors of fields, such classes of large mass interferometers have also been considered as detectors for elusive particles~\cite{kilian_requirements_2023, riedel_decoherence_2017, Du2022atom}. A possible advantage of quantum superposition sensors over classical sensors is the prospect of enhanced sensitivity, since quantum superposition sensors, harnessing the inherent quantum nature of the sensing object, may surpass the SQL.

\subsection{Detection through decoherence}
Levitated quantum sensors are subject to many different types of interactions with the environment. One such mechanism that explains the loss of information due to the localisation of a quantum object as a consequence of these interactions is the \textbf{process of decoherence}. To describe the influence of the environment on the temporal evolution of the sensing object, master equation approaches are a convenient formalism. Instead of having to compute the reduced system's dynamics from the global unitary evolution $\hat{U}(t)$ of the combined system $\hat{\rho}_{S\mathcal{E}}\,(t)$ consisting of the environment and the sensing object, the temporal change in the reduced system's density matrix is generated by a dynamical map $\hat{V}(t)$, yielding an equation of the form $\hat{\rho}_{S}\,(t)=\hat{V}(t) \hat{\rho}_{S}\,(0)$.
For exact master equations, there is a direct correspondence between this dynamical map and the trace over the combined system $\hat{\rho}_{S\mathcal{E}}\,(t)$ and 
\begin{align}
    \hat{\rho}_{S} (t)=\hat{V}(t) \hat{\rho}_{S} (0)\equiv Tr\{\hat{U}(t)\hat{\rho}_{S\mathcal{E}} (t) \hat{U}^\dagger(t)\}.
\end{align}
In general, however, the equation is usually not in a form that would allow for simple calculations and approximate treatments can be used~\cite{schlosshauer_decoherence_2007}. In many situations, what is known as the Born-Markov approximation constitutes a successful model for the dynamical evolution of $\hat{\rho}_{S}(t)$. It is a time-local model, meaning that variations in $\hat{\rho}_{S}(t)$ computed at time $t$ are independent of the behaviour of the reduced system at earlier times $t'<t$. In the Lindblad form, such a master equation reads 
\begin{align}
\label{eq:mastereq_kappa}
\frac{d }{dt}\hat{\rho}_{S} (t)&=-i[\hat{H}_S,\hat{\rho}_{S} (t)]-\frac{1}{2}\sum_{a}\kappa_{a}\{\hat{L}_a \hat{L}^\dagger_a\hat{\rho}_S(t)\\ \nonumber&+\hat{\rho}_S(t)\hat{L}^\dagger_a\hat{L}_a -2\hat{L}_a\hat{\rho}_S(t)\hat{L}^\dagger_a\}
\end{align}
where $\kappa_a$ is the Lindblad coefficient that effectively constitutes a decoherence factor and $\hat{L}^\dagger_a$ are Lindblad operators constructed from combinations of the system operators $\hat{S}_a$. Depending on the physical system, one has to specify the operator $\hat{L}_a$ and the coefficient $\kappa_a$.

Among the principal mechanisms of decoherence in an optomechanical experiment is collisional decoherence, such as decoherence due to collisions with gaseous particles in the environment of the quantum sensor. This type of decoherence can be suppressed to a high degree by increasing the quality of the vacuum, yet it nonetheless constitutes a background for dark matter searches. Its mathematical framework was first developed by Joos and Zeh~\cite{joos_emergence_1985} and later extended by Gallis and Fleming~\cite{gallis_environmental_1990} and Hornberger and Sipe~\cite{hornberger_2003}. Applying equation \eqref{eq:mastereq_kappa} to this physical situation, the equation describing the process of decoherence due to scattering with the bath particles is
\begin{align}
\label{eq:GF}
    \frac{d\rho_S(\mathbf{x},\mathbf{x}')}{dt}&=\frac{1}{i\hbar}\langle \mathbf{x}|[\hat{H}_S,\hat{\rho}_S]|\mathbf{x}'\rangle\\ \nonumber
    &-F(\mathbf{x}-\mathbf{x}')\rho_S(\mathbf{x},\mathbf{x}'),
\end{align}
where $\rho_S(\mathbf{x},\mathbf{x}')$ is the reduced density matrix of the levitated quantum sensor in the position basis, $\hat{H}_S$ is the system Hamiltonian and $F(\mathbf{x}-\mathbf{x}')$ is the localisation rate given by 

\begin{align}
\label{eq:localization_rate}
    F(\mathbf{x}-\mathbf{x}')&=\int dq \ n(q)v(q)\int \frac{d\Omega d\Omega'}{4\pi}\\ \nonumber
    &\times(1-e^{i(\mathbf{q}-\mathbf{q}')(\mathbf{x}-\mathbf{x}')})|f(\mathbf{q},\mathbf{q}')|^2. 
\end{align}

The parameters defining collisional decoherence are the number density distribution of the environment particles $n(q)$ and their speed distribution $v(q)$, the scattering solid angles $\Omega$ and $\Omega'$, the wavevectors $\mathbf{q}$ and $\mathbf{q'}$ of ingoing and outgoing particles and the separation of locations of the quantum sensor $\Delta \mathbf{x}=|\mathbf{x}-\mathbf{x'}|$ (see Figure~\ref{fig:SG-Interferometer}). Depending on the type of scattering, the differential cross-section $|f(\mathbf{q},\mathbf{q}')|^2$ is a measure of the scale and probability of the interaction. Regarding the master equation of particles scattering isotropically in the limit of short scatterer wavelengths $\lambda\ll\Delta \mathbf{x}$, the localisation rate is determined by the total integrated cross-section. Hence, the evolution of the off-diagonal elements of the sensor's density matrix is described by
\begin{align}    \rho_S(\mathbf{x},\mathbf{x}',t)=\rho_S(\mathbf{x},\mathbf{x}',0)e^{-\gamma t}
\end{align}

where $\gamma=\int dq \ n(q)v(q)\sigma_{tot}$ is the scattering constant.

In the long wavelength regime $\lambda\gg\Delta \mathbf{x}$, following a Taylor expansion of the exponent, the localisation rate becomes equivalent to a second order term $\Gamma=\Lambda(\mathbf{x}-\mathbf{x}')^2$, where $\Lambda=\int dq \ n(q)v(q)\int \frac{d\Omega d\Omega'}{4\pi}q^2\sigma_{eff}(q)$. The loss of coherence over time is then expressed through a similar exponential decay
\begin{align} \rho_S(\mathbf{x},\mathbf{x}',t)=\rho_S(\mathbf{x},\mathbf{x}',0)e^{-\Gamma t}.
\end{align}

Note that the scattering constant determines the rate at which the loss of coherence over a time scale $\tau$ manifests. With the basic framework in place, it is possible to estimate the decoherence due to air molecules of average velocity $v$, radius $r$ and mass $m$ interacting with the levitated sensor at a pressure $p$. In the long wavelength regime (LWR), \mbox{$\Lambda_{\text{air,LWR}}=8\sqrt{2\pi}mvpr^2/(3\sqrt{3}\hbar^2)$.} Whereas in the short wavelength regime (SWR), $\gamma_{\text{air,SWR}}=16\pi\sqrt{2\pi}pr^2/(\sqrt{3}mv)$. In the same manner, levitated sensors are also affected by the emission and absorption of thermal photons~\cite{romero-isart_quantum_2011} and other backgrounds.

Particle dark matter, to which one can assign a de Broglie wavelength of 
\begin{align}
    \lambda_\chi&=\frac{2\pi\hbar}{m_\chi\bar{v}_{\chi}},
\end{align}
where $\bar{v}_{\chi}\sim 220\,\text{km/s}$ is the average dark matter velocity, may equally induce the decay of coherences when a candidate scatters from a superposed quantum object. A matter-wave interferometry experiment using silica particles has been proposed~\cite{bateman_existence_2015} for \textbf{searches of light dark matter candidates} with masses $m_\chi\sim100\,\text{eV}$ and nucleus cross-sections on the order of $\sigma_n\sim10^{-29}\,\text{cm}^2$, geared toward measuring the hypothetical particle via its projected decoherence signature. A less restrictive approach to estimating dark matter decoherence rates assumes the scenario of dark matter collisions mediated by an unknown force, modelled by considering Yukawa-type interactions with a differential cross-section

\begin{align}
    \frac{d\sigma_{Y}}{d\Omega}=\frac{g_{\chi}^2g_{M}^2m_{\chi}^2}{4\pi^2(q^2+m_M^2)^2}
\end{align}

where $g_{M}$ and $g_{\chi}$ are the matter and dark matter couplings, $m_M$ is the mediator mass and $m_\chi$ is the dark matter mass. This has been studied~\cite{riedel_decoherence_2017} assuming a spin-independent scattering process for candidate masses ranging from $1\,\text{keV}\, <\,m_{\chi}\,<\,10\,\text{MeV}$, implying deBroglie wavelengths of $1\mu\text{m}>\lambda_\chi>0.1\text{nm}$, and mediator masses $10\,\text{meV}\,<\,m_M\,<\,10\,\text{keV}$. The corresponding decoherence rate that can be derived from this model is then 
\begin{align}
\gamma_{DM}&=\int dq \,n(q)v(q)\\ \nonumber
&\times\int d\Omega\,(1-e^{i\Delta\mathbf{q}(\mathbf{x}-\mathbf{x}')})I(\Delta\mathbf{q})\frac{d\sigma_{Y}}{d\Omega}
\label{eq:dm_decoh_rate}
\end{align}
where $I(\Delta\mathbf{q})$ is a structure factor that depends on the magnitude of $\Delta\mathbf{q}=\mathbf{q}-\mathbf{q}'$ in relation to the target size. If the target size, i.e. the size of a levitated nanoparticle, is smaller than the wavelength $\lambda_{\Delta\mathbf{q}}=2\pi/\Delta\mathbf{q}$ associated with the momentum transfer, then
\begin{align}
I(\Delta\mathbf{q})=\bigg\langle \bigg|\sum_{i=1}^{N}e^{-i\Delta\mathbf{q}\mathbf{x}_i}\bigg|^2\bigg\rangle
\end{align}
where $\mathbf{x}_i$ are the locations of individual nucleons. Hence, if the wavelength spans the full spatial length scale of the regime occupied by the nucleons in the target material, $\Delta\mathbf{q}$ becomes small and an additional enhancement factor to the cross-section scaling with the square of the number of nucleons $N^2$ is obtained, improving the interaction probability. If $\lambda_{\Delta\mathbf{q}}$ is smaller on the other hand, $I(\Delta\mathbf{q})$ will scale only linearly with $N$ but may include an additional contribution in the intermediate regime where an incoming particle may scatter coherently over a reduced volume. Dark matter candidates scattering from volumes superposed over a distance $\Delta\mathbf{x}\ll\lambda_\chi$ will however contribute less to the rate of decoherence, which can be seen upon expansion of the exponential function in Eq.~\eqref{eq:dm_decoh_rate} to second order. The decoherence rate under this expansion is $\Gamma_{DM}=\int dq \,n(q)v(q)\int d\Omega\,(1-\Delta\mathbf{q}^2(\mathbf{x}-\mathbf{x}')^2)I(\Delta\mathbf{q})\frac{d\sigma_{Y}}{d\Omega}$. 
Provided the decoherence rates due to gas collisions and black body radiation can be sufficiently suppressed, the excess decoherence signature expected from particulate dark matter events may be distinguishable from the typical experimental background. However, the scattering of neutrinos will constitute one such irreducible background which may be difficult to distinguish from a dark matter event in any realistic experimental setup and in particular in the low MeV energy regime, where neutrinos will interact with matter predominantly via neutral current coherent-elastic neutrino-nucleus scattering~\cite{lindner_coherent_2017,kilian_requirements_2023}.

\begin{table*}
    \centering
    \begin{tabular}{c|c|c|c|c}
        Levitated & 
        \multicolumn{2}{c}{Mechanical Sensitivity} & 
        Testable DM &
        DM Parameter Space probed \\
        System &
        $\sqrt{S_a}\,(g/\sqrt{\text{Hz}})$ & 
        $\sqrt{S_F}\,(\unit{N/\sqrt{\text{Hz}}})$ &
        Candidates &
        [DM Mass range] 
        \\
        
        \hline
        
        Optically &
        $\sim\,\num{6d-6}$\,--\,$\num{9d-8}$ &
        $\sim\,\num{1d-18}$\,--\,$\num{6d-21}$ &
        Millicharge &
        Charge\,[e]\,$\sim$\,$10^{-4}$\,[GeV\,--\,TeV]$\dagger$~\cite{afek2021limits} \\ 
        trapped & 
        &&
        Composite & 
        $\sigma_{\chi n}$\,[cm$^2$]$\,\sim\,10^{-28}$\,[$10^3$\,--\,10$^4$\,GeV]$\dagger$~\cite{monteiro_search_2020} \\ 
        (fg-ng)~\cite{liang2023yoctonewton,monteiro2020force} &
        &&
        Low-mass particle &
        $\sigma_{SI}$\,[cm$^2$]$\,\sim\,10^{-30}$\,[0.1\,--\,100\,MeV]*~\cite{afek2022coherent} \\ 
        &&&
        Sterile $\nu$ &
        $\left |U_{e4} \right |^2\,\sim\,10^{-4}\,$--$ 10^{-6}$\,[0.1\,--\,1\,MeV]*~\cite{carney2023searches} \\ 
        \hline
        
        Magnetically &
        $\sim \,\num{1d-10}\,$\,--\,$\,\num{9d-12}$ &
        $\sim \,\num{5d-12}\,$\,--\,$\,\num{5d-19}$ &
        ALPs &
        $g_{aee}\,\sim$ $10^{-14}$\,[$10^{-13}$ to $10^{-18}$\,eV]~\cite{ahrens2024levitated} \\ 
        trapped &
        &&
        Axions &
        $g_{a\gamma}\,[GeV^{-1}]\,\sim\,10^{-10}\,[10^{-12}$\,--\,$10^{-14}$\,eV]~\cite{higgins2023maglev} \\ 
        ($\mu$g-mg)~\cite{hofer_highq_2023,timberlake2019acceleration} &
        &&
        Dark photons &
        $\epsilon\,\sim\,10^{-8}\,[10^{-12}$\,--\,$10^{-14}$\,eV]~\cite{higgins2023maglev} \\ 
        &&&
        ULDM &
        $g_{B-L}\,\sim\,10^{-25}$[$10^{-14}$ to $10^{-16}$\,eV]~\cite{li2023search} \\ 
        \hline
        
        Electrically &
        $\sim \,\num{5d-6}$ &
        $\sim \,\num{2d-21}$ &
        ULDM~\cite{budker2022millicharged} & no concrete experimental proposals
        \\
        trapped &
        &
        &
        Composite& beyond trapped atomic ions\\
        (40 fg)~\cite{dania2023ultrahigh} &&&&\\
        \hline
    \end{tabular}
    \caption{Table showcasing the mechanical sensitivities for the different types of levitation ($\sqrt{S_a}$ is the acceleration sensitivity; $\sqrt{S_F}$ the force sensitivity). Also shown are theoretical and experimental sensitivities to different dark matter candidates and the parameter space explored therein. The square brackets indicate the mass ranges explored. The symbol * represents proposals utilising multiple trapped particles, while $\dagger$ indicates experimentally measured sensitivities.}
    \label{tab:DMtable}
\end{table*}
\subsection{Coherent detection through quantum mechanical phases}

The primary focus of detecting particulate matter through matter wave interferometry has in recent years been placed on sensing the environment via the decoherence produced due to an interaction. The particles hitting the sensor are assumed to scatter with varied momenta and impinge from all directions, which is mathematically represented by averaging over incoming scattering angles. The effect of this averaging is an elimination of the imaginary contribution to the localisation parameter. In the short wavelength limit, the rapid oscillatory behaviour of the exponential which forms part of the integrand in Eq.~(\ref{eq:localization_rate}) averages out upon integration. The long wavelength limit allows for a Taylor expansion of the integrand which equally results in the disappearance of the imaginary contribution~\cite{schlosshauer_decoherence_2007}.
For a \textbf{directional source} such as a stream of dark matter particles with a narrow distribution of incoming angles, however, there is also the possibility that a scattering of the particle with the sensor may result in the \textbf{manifestation of a coherent contribution} to the phase $\Delta\phi$ measured at the end of an interferometry experiment~\cite{kilian_optimal_2023}. In this case, the imaginary part would not vanish and may even be more pronounced than the decoherence produced by the scattering. Let us exemplify this statement by comparing the first and second-order terms of the series expansion in $\Delta x$ in the long wavelength limit for small momentum transfers. Up to constants of proportionality, the expansion of the exponential to second order in $\Delta x$ will yield terms $iq\Delta x$ and $q^2\Delta x^2$. This implies that if the ratio $q\Delta x\ll 1$, the first-order term may potentially dominate over the second-order decoherence contribution. Depending on the characteristics of the particle source and the choice of the superposition size $\Delta x$, the effect may persist after performing all integrations and result in a coherent phase contribution. The evolution of the initial quantum state of the sensing object 
\begin{align}   \rho_S(\mathbf{x},\mathbf{x}',t)=\rho_S(\mathbf{x},\mathbf{x}',0)e^{-\Gamma+i\delta\phi}
\end{align}
will then not only be characterised through a real-valued scattering decoherence parameter $\Gamma$, but also through a relative phase shift $\Delta\phi$ accumulating between superposed components. It should be noted however that in order for the phase effects to be measurable, the visibility must be sufficiently large, implying that the decoherence is required to be low. The decoherence and phase effects due to dark matter scattering events have recently~\cite{Du2022atom} been analysed in more detail for atomic as well as large-mass interferometers and dark matter masses in the range $10\lesssim m_{\chi}\lesssim 10^5\,\text{eV}$. The considered interaction channels include nuclear recoil, coherent axion scattering and hidden photon processes. The mathematical modelling suggests that interferometric approaches for searches in the aforementioned light dark matter mass regime could improve current astrophysical bounds by an order of magnitude.
 
\section{Summary}

In summary, we have presented a perspective on the prospect of employing levitated sensors for dark matter searches. Following an overview of select classes of dark matter models, we discussed the possibility of levitated free space and cavity-based large-mass systems being used for the detection of both ultralight and hidden-sector candidates, emphasising the potential sensitivity enhancement of recently introduced cross-correlation techniques. We further highlighted theoretical proposals for future interferometric techniques, based on superpositions of massive levitated objects, for sensing certain types of dark matter through their scattering decoherence signatures as well as coherent phases accumulating between superposed components.

Table ~\ref{tab:DMtable} summarises the already sizable body of work in the literature, which includes both experimental results setting new limits, and projections for the sensitivity of future experiments. For instance, the search for hypothesised millicharged dark matter carrying electrical charge as conducted~\cite{afek2021limits} with optically trapped nanoparticles has set new limits on their abundance by probing the mass to fractional charge parameter space. Capable of searching for electric charges as small as $10^{-4}\text{e}$, this experimental approach is uniquely suited to explore regions in the parameter space beyond the scope of present large-scale underground and accelerator experiments. This provides a strong motivation for further development of levitated sensors for dark matter searches. A similar experiment~\cite{monteiro_search_2020} looking for impulse-like signatures of composite dark matter candidates in the TeV mass regime has also placed novel limits on dark matter-neutron interactions and dramatically exceeded the sensitivity of prior searches.

Proposed future experiments for directional searches using optically trapped femtogram masses (or possibly arrays thereof) that exploit large enhancements in the cross-section for spin-independent scattering due to coherent scattering of the dark matter candidates from the entire sensor, are expected to show increased sensitivity to dark matter masses below 1\,GeV~\cite{afek_coherent_2022}. Approaches capable of full momentum reconstruction may also be used to conduct searches for sterile neutrinos~\cite{carney2023searches}, possibly yielding novel limits to the mixing matrix element $|U_{e4}|^2$, an extended form of the famed Pontecorvo-Maki-Nakagawa-Sakata matrix, placing new constraints in the neutrino mass range $0.1 - 1 \ \text{MeV}$.

Magnetically trapped micron-sized ferrimagnets have been proposed for probes of axion dark matter in the galactic halo model~\cite{ahrens2024levitated} in the mass regime \mbox{$10^{-18}\,-\,10^{-13}\,\text{eV}$}. Although current magnetometers are not yet able to surpass the bounds set by existing astrophysical observations and solar axion searches~\cite{aprile2022search} for the coupling of axions to electrons, $g_{aee}$, the use of larger magnets is expected to offer comparable and potentially increased sensitivity. Levitated superconducting particles~\cite{higgins2023maglev} have been studied for ultralight dark matter detection in mass regimes $\num{4d-15}\,\lesssim m_{\chi}\,\lesssim\,\num{4d-12}\,\text{eV}$ and they may be employed to constrain the dark photon coupling $\epsilon$ and the axion-photon coupling $g_{a\gamma}$, with a predicted achievable sensitivity comparable to other laboratory probes. Diamagnetic levitated sensors with a changeable resonance frequency have been suggested for ultralight dark matter detection in a similar mass range of $10^{-16}\,\lesssim\,m_{\chi}\,\lesssim\,10^{-13}\,\text{eV}$ for dark matter vector fields coupling to the standard model through baryon-minus-lepton number channels~\cite{li2023search}, claiming a possible tenfold improvement in $g_{B-L}$ limits.

Given the already experimentally demonstrated successes of levitated setups to place new constraints on dark matter parameter spaces as well as the exciting possibilities outlined by proposals for future experiments, quantum technologies based on large-mass levitated systems are able to probe regions beyond the capabilities of present-day large-scale experiments and offer strong complementary approaches to other dark matter searches.

\section*{Acknowledgements}
EK would like to acknowledge support from the Engineering and Physical Sciences Research Council (grant number EP/L015242/1) and an EPSRC Doctoral Prize Fellowship Grant (EP/W524335/1). MT would like to acknowledge funding by the Leverhulme Trust (RPG- 2020-197). SB would like to acknowledge EPSRC grants (EP/N031105/1 and EP/S000267/1) and grant ST/W006227/1. MR, JMHG, FA, AP and PFB acknowledge funding from the EPSRC and STFC via Grant Nos. EP/N031105/1, EP/S000267/1, EP/W029626/1, EP/S021582/1 and ST/W006170/1. CG would like to acknowledge STFC grant no. ST/W006227/1 and additional support from the Cosmoparticle Initiative. JHI and TSM acknowledge funding from the EPSRC Grant No. EP/W029626/1. 

\section*{Author Declarations}
\subsection*{Conflict of interest}
The authors have no conflicts to disclose.

\section*{Data Availability}
Data sharing is not applicable to this article as no new data were created or analyzed in this study.

\bibliographystyle{apsrev4-1} 

\bibliography{references.bib}

\end{document}